\pdfoutput=1
\documentclass{jfm}
\usepackage{graphicx}
\usepackage{epstopdf, epsfig}
\usepackage{xcolor}

\shorttitle{Critical point in pipe flow}
\shortauthor{V. Mukund and B. Hof}

\title{The critical point of the transition to turbulence in pipe flow}

\author{Vasudevan Mukund\aff{1}
\and Bj{\"o}rn Hof\aff{1}
\corresp{\email{bhof@ist.ac.at}},}
\affiliation{\aff{1}Institute of Science and Technology Austria, Am Campus 1, 3400 Klosterneuburg, Austria}

\graphicspath{{./figs/}}

\begin{document}

\maketitle

\begin{abstract}

In pipes, turbulence sets in despite the linear stability of the laminar 
Hagen-Poiseuille flow. The Reynolds number ($Re$) for which turbulence 
first appears in a given experiment - the `natural transition point'- 
depends on imperfections of the set-up, or more precisely, on the 
magnitude of finite amplitude perturbations. At onset, turbulence 
typically only occupies a certain fraction of the flow and this fraction 
equally is found to differ from experiment to experiment. Despite these 
findings, Reynolds proposed that after sufficiently long times, flows may 
settle to a steady condition: below a critical velocity flows should 
(regardless of initial conditions) always return to laminar while above 
eddying motion should persist. As will be shown, even in pipes several 
thousand diameters long the spatio-temporal intermittent flow patterns 
observed at the end of the pipe strongly depend on the initial 
conditions and there is no indication that different flow patterns 
would eventually settle to a (statistical) steady state. Exploiting the 
fact that turbulent puffs do not age (i.e. they are memoryless), we 
continuously recreate the puff sequence exiting the pipe at the pipe 
entrance, and in doing so introduce periodic boundary conditions for the 
puff pattern.

This procedure allows us to study the evolution of the flow patterns for 
arbitrary long times, and we find that after times in excess of $10^7$ 
advective time units, indeed a statistical steady state is reached. Though the 
resulting flows remain spatio-temporally intermittent, puff 
splitting and decay rates eventually reach a balance so that the 
turbulent fraction fluctuates around a well defined level which only 
depends on $Re$. In accordance with Reynolds proposition, we find that at 
lower $Re$ (here 2020), flows eventually always resume to 
laminar while for higher $Re$ ($>=2060$) turbulence persists.

The critical point for pipe flow hence falls into the interval of 
$2020<Re<2060$, which is in very good agreement with the recently 
proposed value of $Re_c=2040$. The latter estimate was based on single puff statistics and entirely neglected puff interactions.

Unlike in typical contact processes where such interactions strongly affect the percolation threshold, in pipe flow the critical point is 
only marginally influenced. Interactions on the other hand, 
are responsible for the approach to the statistical steady state. As shown, they strongly affect the resulting flow patterns, where they cause `puff clustering' and these regions of large puff densities are observed to travel across the puff pattern in a wave like fashion.

\end{abstract}


\section{Introduction}
\cite{reynolds1883experimental}  introduced the dimensionless ratio $Re=UD/\nu$ ($U$ being the mean or bulk flow speed, $D$ the pipe diameter and $\nu$ the fluid's kinematic viscosity), now called the Reynolds number, as the sole parameter that governs the flow through a straight pipe. Here, the bulk speed is given by $ U = Q/A $, where Q is the flow-rate and A is the cross-sectional area of the pipe. In the case of laminar flow, $U = u_{c}/2$, where $u_{c}$ is the center-line velocity. Reynolds observed that, at the onset, turbulence took the form of localized patches (`flashes') surrounded by laminar flow. These are now commonly referred to as `puffs' and localized, non-expanding puffs are  known to occur at   $ 1700 \lesssim Re \lesssim 2300$. He also observed that depending on the care with which the pipe was set up, the Reynolds number at which turbulence was observed could be postponed from  $Re \approx 2000$ to  $Re \approx 13000$. Reynolds correctly concluded that the reason for the variation of this natural transition point is the sensitivity of the flow to perturbations of finite amplitude (and stability to infinitesimal ones) or in his words \cite[p.956]{reynolds1883experimental}: `This at once suggested the idea that the condition might be one of instability for disturbance of a certain magnitude and stable for smaller disturbances'. 

It is now believed that Hagen-Poiseuille flow is linearly stable for all $Re$ \citep{meseguer2003linearized}, and that turbulence can only be triggered by finite amplitude perturbations. Hence, the Reynolds number where turbulence naturally occurs in a given setup is not universal. Nevertheless, Reynolds proposed the existence of a `real critical value' which is the same for all pipes: `it became clear to me that if in a tube of sufficient length the water were at first admitted in a high state of disturbance, then as the water proceeded along the tube the disturbance would settle down into a steady condition, which condition would be one of eddies or steady motion, according to whether the velocity was above or below what may be called the real critical value.' \cite[p.958]{reynolds1883experimental}. In addition to a critical point, he proposes a steady state which he presumes to be laminar below critical and consisting of eddies (i.e. turbulent) above critical. While many propositions for the critical point have been made in the past, the approach to a (statistical) steady state could not be demonstrated so far.

Although the above definition of the critical point may seem straightforward, Reynolds could not obtain an exact value. In his initial study, he assumed the value to be close to 2000 while in his subsequent investigation \citep{reynolds1894dynamical} he proposed the value to be somewhere in the range of 1900 to 2000. 
In later experiments, \cite{schiller1921} reported a discontinuous change in friction factor at $Re=2320$ and proposed this as the critical point. \cite{binnie1945} and \cite{binniefowler} used visualization methods and observed turbulence to first appear at $Re$ ranging from 1900 to 1970; the precise value appeared to depend on the turbulence levels at the inlet. \cite{lindgren1953some} found that, in contrast to Schiller's observations \citep{schiller1921}, the onset of intermittent turbulence did not coincide with a discontinuity of the friction factor, and argued that the critical point could not be determined by friction factor measurements. \cite{rotta1956} attempted to determine the critical value by measuring growth rates of turbulent regions. He found that for $Re < 2300$, the growth rates were too slow to be accurately determined within the length of the pipe, but he suggested a value close to 2000 by extrapolating growth rates down to zero. \cite{stern1970} found values between 2300 and 2400, appearing to depend on the pipe diameter.
\cite{sibulkin} studied the behavior of initially turbulent flow with different inlet conditions. He found that the critical point was between 1800 and 2400, though, even at those Reynolds numbers, the pipe was too short for the flow to reach a statistical steady state. \cite{pavelyev2003} and \cite{pavelyev2006} found that  the Reynolds number at which disturbances decayed depended on the intensity and nature of the perturbations. They suggest this dependence to be the cause of the wide scatter in the critical value obtained in previous studies. 

Other difficulties in determining the critical point more accurately are the advective nature of turbulence (in this $Re$ regime, turbulent structures travel downstream at approximately the bulk speed) and the fact that turbulence, even if it has been triggered successfully, can disappear again at later times \citep{brosa1989turbulence}. A detailed investigation of the latter effect carried out in numerical simulations of a  $5 D$ pipe with periodic boundary conditions, determined that the decay of turbulence follows a memoryless process, and hence that turbulent structures do not age but rather decay suddenly \citep{Faisst2004}. Detailed lifetime statistics revealed a steep increase of lifetimes with $Re$ and on this basis a critical point of $2250$ was proposed where lifetimes become infinite. 
Later lifetime studies focused on the behaviour of actual puffs, which are considerably larger than the $5 D$ of the above study. In an experimental investigation by \cite{peixinho2006decay}, lifetimes were reported to diverge at $Re$ as low as $1750$ and a subsequent numerical study \citep{willis2007critical}, using a domain size of $50 D$, found $Re_c=1870$. 
However studies in much longer pipes and based on observations of much larger numbers of puffs \citep{Hof2006a,Hof2008,DeLozar2009,avila2010transient}, found on the contrary that lifetimes do not diverge, and speculated that turbulence may be intrinsically of transient nature. Although puff lifetimes did increase with Reynolds number, they remained finite, and no critical number could be determined when considering decay of spatially isolated puffs.

An alternative way to determine the critical point, taking spatial proliferation into account was proposed by \cite{avila2011onset}. Here, in addition to the decay rates of isolated turbulent puffs, the growth rate associated with puffs was also determined statistically. The growth process is commonly referred to as puff splitting. Puffs shed vortices from their leading edge, which typically decay. Occasionally however, such vortices can persist and separate sufficiently far from the original puff to seed a new one. While the lifetimes of puffs increase rapidly with $Re$, the time before a splitting happens decreases with $Re$. The point where the two processes balance, i.e. where new puffs are created at the same rate as individual ones decay, was found to occur at $Re_{c} \approx 2040$ and proposed as the critical $Re$ above which turbulence becomes sustained. At this $Re$, the characteristic times for decay and splitting are equal to approximately $2 \times 10^{7} D/U$ (advective time units). Since both processes vary super-exponentially with $Re$, only slightly above the balance point, the creation of new puffs will sufficiently outweigh puff decay, and it is hence plausible to assume that here turbulence is likely to be sustained. However this estimate does not take interactions and the complexity of the spatio-temporally intermittent puff patterns into account. The relevance of such spatial interactions and coupling between chaotic and `laminar' regions have been pointed out previously for coupled chaotic maps (coupled map lattices) by Kaneko 1985. Importantly, due to these spatial interactions, here a new dynamical state `spatio-temporal intermittency' has been identified. The spatial coupling drives a continuously changing coexisting pattern of chaotic and laminar sites. Analogies to laminar-turbulent patterns in shear flows have been pointed out subsequently \citep{bottin1998discontinuous,bottin1998statistical,manneville2014transition,moxey2010distinct}. All these studies imply that spatio-temporal intermittency is intrinsic to the transition to turbulence in a variety of linearly stable flows. Earlier studies in pipe flow on the other hand had suggested that the steady state approached above the critical point (`state of eddies' as proposed by
Reynolds) would be uniformly turbulent \citep{rotta1956}. 

Generally, interactions also affect the critical point which can be readily seen for basic non-equilibrium contact processes like directed percolation (DP). Here the criterion for criticality chosen by \cite{avila2011onset}, the balance point between spreading and decay, would fail. For directed percolation this would correspond to a probability (the control parameter in percolation) $P=0.5$.
 Instead (due to interactions) the actual critical point is only reached at $\approx 30 \%$ higher probabilities ($P_{c}=0.64$). While for pipe flow, the shift between the actual critical point and the balance point is likely to be small due to the super-exponential change in decay and splitting times with $Re$ (a property models of DP do not share), the statistical steady state can only be reached if the actual splitting and decay rates have reached a balance and interactions are crucial for this. The time scales required to reach this equilibrium state close  to the critical point are assumed to exceed splitting and decay times of isolated puffs ($\approx 2 \times 10^{7}$ advective time units).  Hence, due to puff advection, a pipe length of order $10^{8} D$  would be required, making such an undertaking seemingly impossible in practice.
 
By exploiting the memoryless nature of turbulent puffs, we implement effectively periodic boundary conditions in  experiments. In doing so we can achieve observation times many orders of magnitude longer than in any previous study. As shown, just above the critical point, spatio-temporal intermittency persists and a statistical steady state is approached where puff creations and decays are in balance. While the critical point appears to be hardly affected, interactions give rise to puff clustering and correlations on length scales longer than previously expected.
\section{Pipe experiments}
\label{setup}

\begin{figure}
 \includegraphics[scale=0.4]{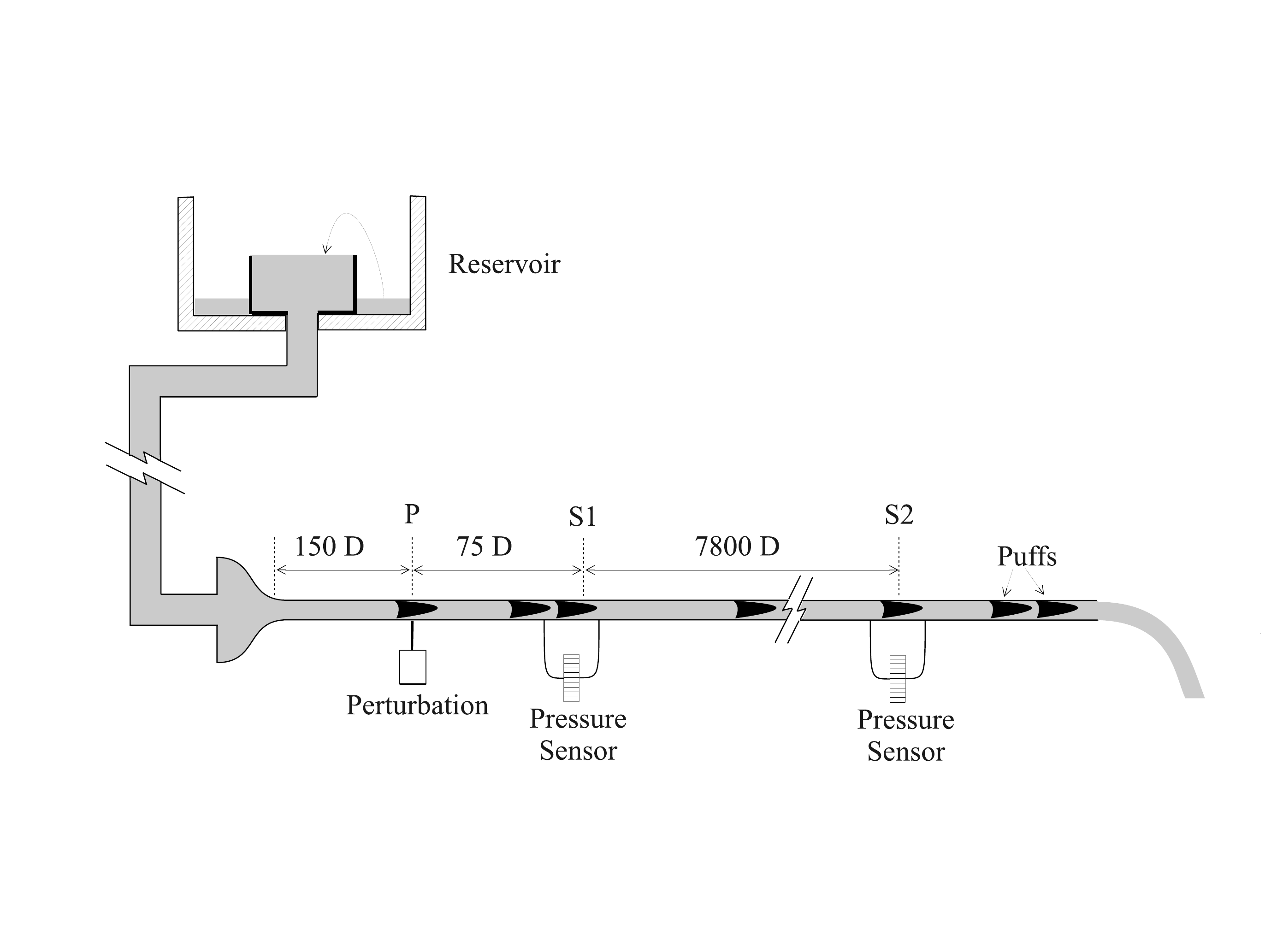}%
 \caption{A schematic of the gravity driven pipe set-up.} 
 \label{fig:schematic}
 \end{figure}

Experiments are carried out in two different pipe setups. In the first setup, the pipe was made of glass tubes with an inner diameter of $D = 2 $ mm $(\pm 0.007$ mm), and has a total length of $8000 D$. Pipe sections are joined with 4 cm long acrylic connectors \citep{samanta2011experimental,DeLozar2009} that allow to accurately match and align the inner diameter of the tubes. Some of the acrylic connectors had 0.4 mm holes drilled into them (perpendicular to the pipe axis) that can be used to perturb the flow or to measure the pressure. The working fluid is water which is supplied from a reservoir (see figure~\ref{fig:schematic}) positioned more than 20 metres above the pipe. The reservoir is continuously overflowing, ensuring a precise pressure head to drive the flow. The reservoir is mounted on motorized, vertical guide-rails, permitting fine adjustments to the height of the reservoir, and hence the flow rate.
The fluid enters the pipe through a convergence which assures laminar flow up to $Re \approx 5000$ (natural transition point).

The flow rate is measured in an automated fashion by collecting the water at the pipe exit over a fixed period (typically 50 seconds, corresponding to around $2.5 \times 10^{4}$ advective time units for the Reynolds number range investigated) and weighing the amount of water with electronic scales, thus obtaining the flow rate at regular intervals with an accuracy of better than $0.2\%$.
The fluid temperature is monitored by platinum RTD probes at the inlet and outlet of the pipe. Throughout the measurements, temperature differences between these probes remained below $0.1$ K, though the absolute temperature could change by several degrees. The mean value of these two probes is taken to be the temperature of the fluid in the pipe. $Re$ number is then determined from the flow rate and taking the temperature dependence of the fluid's viscosity into account.

The Reynolds number can change due to variations in fluid temperature or turbulent fraction (due to splitting and decay of puffs). For instance, the difference in frictional drag due to addition or disappearance of a single puff changes the Reynolds number by around $1$. Any such change in the Reynolds number results in an adjustment of the vertical position of the reservoir to bring $Re$ back to its original value.  Using this feedback method, an overall stability of the Reynolds number to within $\pm 5$ is achieved.
 

The flow is monitored by two differential pressure sensors  (Validyne DP45) one located $75 D$ (point S1 in figure~\ref{fig:schematic}) and the second $7875 D$ (point S2 in figure~\ref{fig:schematic}) downstream of the perturbation point. Each sensor measures the pressure difference across two holes separated by $8 D$  in the streamwise direction.
Turbulent puffs cause a local increase in skin friction and hence their passage causes a distinct increase in the pressure loss across the two points. Thus, the passage of a puff shows up as a peak in an otherwise flat pressure signal (see, for example, figure~\ref{fig:puff reproduction}).


In broad outline, the second pipe set-up is identical to the first one. A crucial difference is that the pipe is of larger diameter ($D = 4 \pm 0.01$ mm), and is shorter, with a length of $2500 D$. The pressure sensors monitor the flow at $250 D$ and $2500 D$. The natural transition for this pipe was lower, being $Re \approx 3000$.

\section{Characteristics of puff turbulence}
\label{puff characteristics}
\begin{figure}
 \includegraphics[scale=0.7]{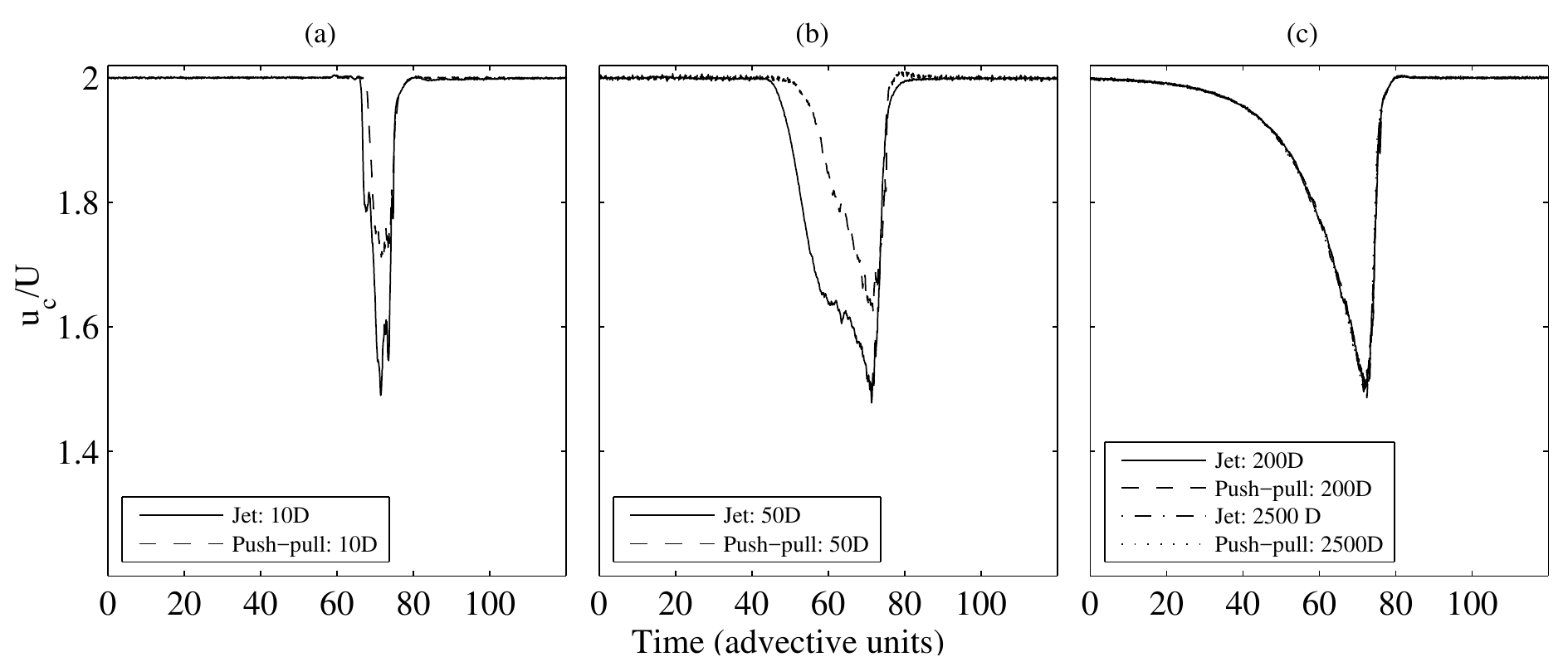}
 \caption{Ensemble averaged centreline velocity of turbulent puffs at different times after the perturbation, for an impulsive jet and a push-pull perturbation. For times smaller than around 200 advective time units, the average velocity is still evolving, and the signals from the two perturbations are different. Beyond this, the signals agree very well, and there is no further evolution in time, emphasizing the memoryless nature of puffs.}
 \label{fig:phaseavg}
 \end{figure}

Turbulence at moderate Reynolds numbers ($1700 \lesssim  Re \lesssim  2300$) in pipes occurs in the form of localized puffs separated by laminar regions. Laminar here refers to vanishing fluctuation levels, though the profile may not have developed its parabolic shape yet. In the interior of a puff, vortices are predominantly produced close to the upstream (trailing) edge \citep{hof2010eliminating}. Vortices are then advected downstream forming an elongated arrow shaped downstream (leading) edge and their amplitude decreases exponentially with distance from the upstream edge. The energetic part of puffs can be taken to be approximately $5 D$ to $10 D$  in length,  while the downstream distance over which the profile recovers its parabolic shape is substantially larger \citep{hof2010eliminating}. Puffs keep their shape and characteristic size for long times before they suddenly decay. Puffs however also interact: If two puffs are separated by less than $25 D$,  the downstream puff will decrease in amplitude and eventually decay. Equally, it has been shown \cite{hof2010eliminating} that flattening the velocity profile directly upstream of a puff results in its decay. Energetically, puffs appear to depend on the upstream laminar flow with a sufficiently parabolic profile \citep{samanta2011experimental}. The region directly downstream of a puff on the other hand is characterized by a flat velocity profile with decaying fluctuation levels and here the flow is `refractory'. Even when external perturbations are applied to this region, they will swiftly decay and no additional turbulence can be sustained here. With further downstream distance, the flow profile slowly recovers its parabolic shape and once the profile is sufficiently developed (at least $25 D$ downstream of the original puff), a second puff can be triggered \citep{samanta2011experimental}. Much of this qualitative behavior can be captured by a reaction-diffusion model \citep{barkley2011simplifying,barkley2015rise} where localised excitations equally require a minimum time during which the system has to recover (refractory region). 

As will be shown in the following, puffs are the only long lived turbulent structures that can be excited at moderate Reynolds numbers ($1700 \lesssim $ Re $\lesssim 2300$). Fully turbulent flow  is unstable in this Reynolds number regime. If triggered, it quickly collapses into a sequence of puffs. Puffs are statistically identical and any structural differences resulting from the initial disturbance creating them is quickly lost.
It is also shown that not all perturbation mechanisms are capable of triggering puffs and that the observed puff frequency (turbulent fraction) immediately downstream of the disturbance point strongly depends on the perturbation used.

The evolution of individual puffs and the dependence of their structure on the initial perturbation is investigated by comparing puffs triggered by two different perturbations in the 4 mm pipe setup. The impulsive jet perturbation involved injecting a jet of water orthogonal to the flow for a short period of time, through a hole of diameter 0.4 mm. The timing was controlled by a solenoid valve. The injection time was approximately $2 D/U$, and the relative mass flux of the perturbation to that in the pipe was around $10^{-2}$. The other perturbation was a push-pull type, where an actuator was used to simultaneously inject and withdraw water from the pipe through two diametrically opposite holes, each of diameter 0.4 mm. The duration of this perturbation was around $20 D/U$ and the relative mass flux was around $10^{-3}$.

The centre-line velocity was monitored at different distances downstream of the perturbation using Laser Doppler Velocimetry (LDV). For each perturbation type, the velocity signals were phase-averaged over 50 puffs and the results are shown in figure~\ref{fig:phaseavg}. As can be seen from the signals at $10 D$ and $50 D$, the perturbations are still evolving, and their shape depends on the perturbation type. However, beyond around $200 D$ from the perturbation point, the puffs have developed their characteristic structure and regardless of the initial perturbation they are statistically identical (here shown for the puff's center-line velocity signature). This is in line with \cite{wygnanski1973transition} who reported developed puffs to be identical and independent of the original disturbance. Equally it has been reported by \cite{DeLozar2009} that puff lifetimes are independent of the initial perturbation. That puffs are statistically indistinguishable and appear to have no memory of how they were created, is rooted in their chaotic properties and the sensitive dependence on initial conditions on chaotic sets which erases memory exponentially fast.

\begin{figure}
 \includegraphics[scale=0.7]{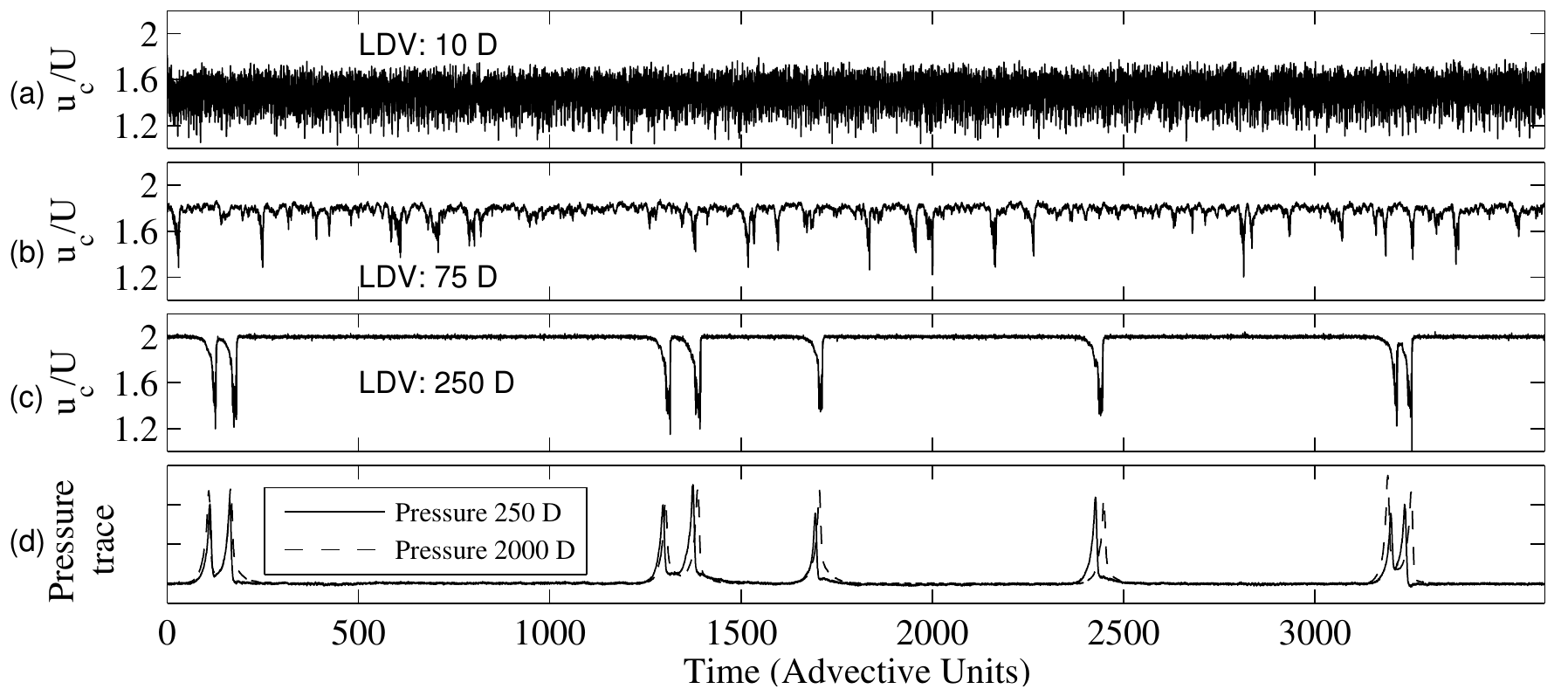}%
 \caption{Evolution of strongly disturbed inlet flow at  $Re = 2060$ in the 4 mm pipe. The top three panels show the ratio of the centre-line velocity $u_{c}$ (obtained by Laser Doppler Velocimetry) to the bulk velocity U at $10 D$, $75 D$  and $250 D$ respectively from the entrance. The bottom panel shows the pressure signals at $250 D$ and $2500 D$.}
 \label{fig:collapse}
 \end{figure}
 
Next, we look at the evolution of a fully turbulent flow. Figure~\ref{fig:collapse} shows the time series of the ratio of centre-line velocity (obtained by LDV) to the bulk velocity, as well as pressure traces at different locations downstream of a turbulent pipe inlet at a Reynolds number of 2060. These measurements are carried out in the 4 mm pipe, where a nozzle with a 2 mm diameter is connected to the pipe inlet continuously perturbing the flow. $10 D$ downstream of the perturbation point (figure~\ref{fig:collapse}a), the flow appears to be uniformly turbulent, with strong velocity fluctuations, while the mean value of the center-line velocity deviates strongly from the case of laminar flow. At $75 D$ from the perturbation point however, the turbulence has broken up into narrow active regions separated by more quiescent flow (figure~\ref{fig:collapse}b). The centre-line velocity shows distinct minima whereas in between, fluctuation levels are lower and the centre-line velocity is closer to the laminar value. As seen in figure~\ref{fig:collapse}c, at $250 D$, the flow has eventually settled to a sequence of distinct puffs separated by regions of laminar flow. At the location of each puff, the centre-line velocity drops sharply well below the laminar value of 2U while the pressure drop (figure~\ref{fig:collapse}d) across a puff is larger (due to the increased drag) and hence shows a distinct peak compared to the surrounding laminar flow. For any perturbation tested in this Reynolds number regime (see more examples below), flows will either settle to a sequence of puffs or relaminarize. 

Although puffs are internally highly chaotic, sequences of puffs appear to be rather stable. Comparing the pressure signal (figure~\ref{fig:collapse}d) at $250 D$ to that recorded at $2500 D$ the puff pattern is practically unchanged. This raises the question whether the puff pattern has reached a steady state or evolves on longer time scales.

\begin{figure}
 \center
 \includegraphics[scale=0.6]{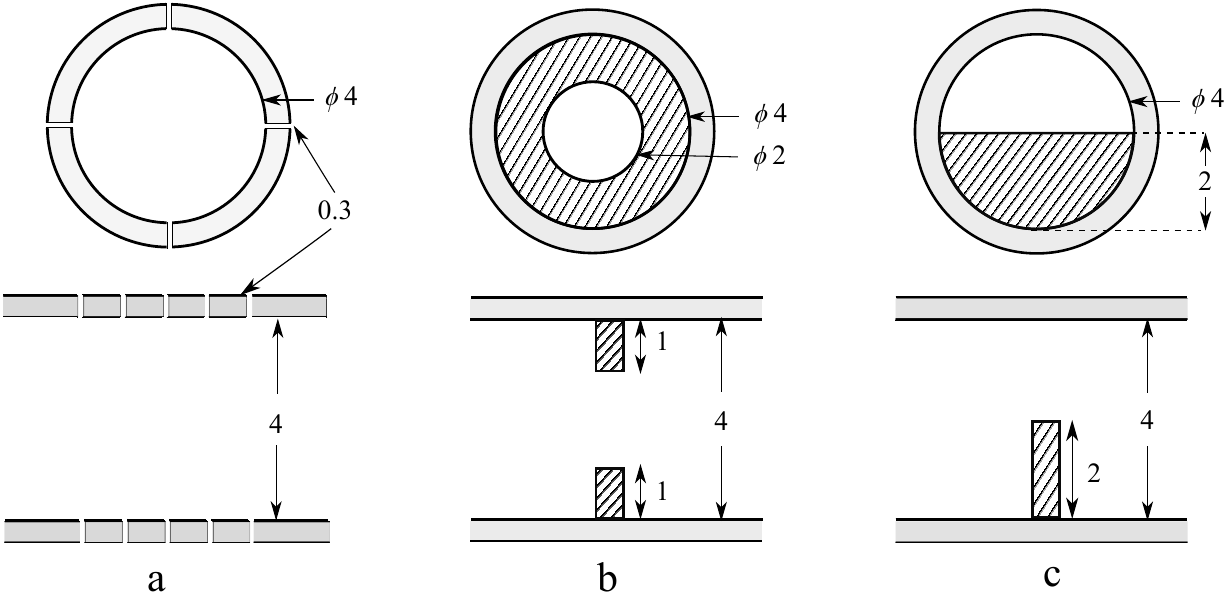}%
 \caption{Radial (top row) and axial (bottom row) cross sections of different disturbance geometries used to perturb the flow: all dimensions are in mm (a) multiple jets: continuous injection through 20 holes in the pipe wall, each having a diameter of 0.4 mm, (b) orifice with an open diameter of 2 mm, and (c) semi-circular obstacle with a diameter of 4 mm. }
 \label{fig:PertSchematic}
 
 \vspace*{\floatsep}

 \centerline{\includegraphics[scale=0.7]{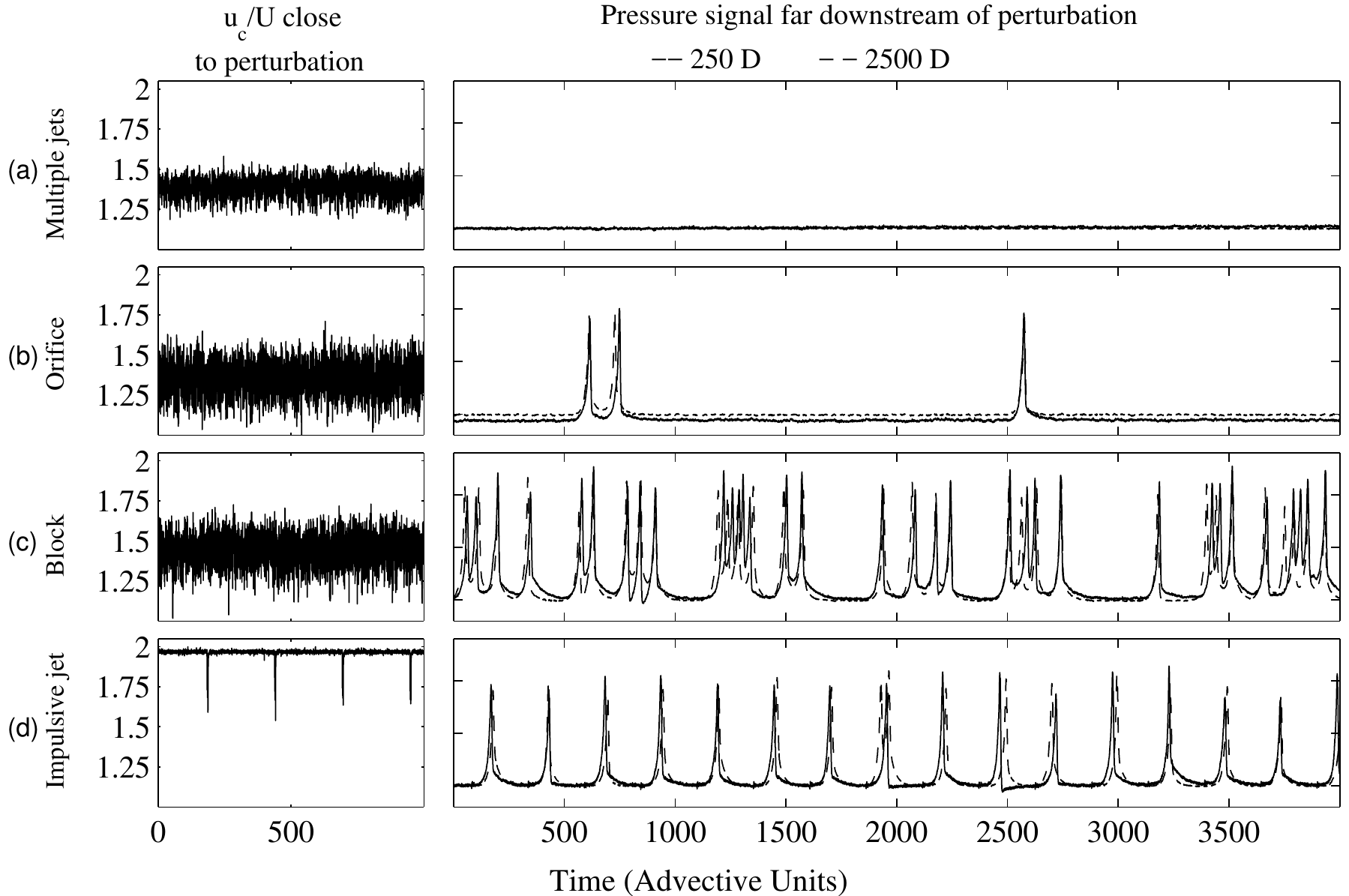}}%
 \caption{Effectiveness of different perturbations in exciting puffs at a Reynolds number of 2100. The sub-panels on the left show the ratio of the centre-line velocity $u_{c}$ (measured by LDV)  to that of the bulk flow velocity $U$, $10 D$ downstream of the perturbation. The sub-panels on the right show the corresponding pressure signals $250 D$ and $2500 D$ downstream of the perturbation. (a) multiple jets: continuous injection through 20 holes in the pipe wall (b) 2 mm orifice (c) semi-circular obstacle (4 mm diameter) (d) impulsive jet perturbation (injection time of $\approx  2 D/U$) }
 \label{fig:obstacles}
 \end{figure} 
 
While puffs can be excited at Reynolds numbers as low as 1700, not all perturbation mechanisms are suitable to trigger puffs, especially at low $Re$. This is illustrated by measurements carried out in the 4 mm pipe at $Re = 2100$ using different perturbations. In addition to the single jet perturbation already described earlier in this section, three other perturbation mechanisms are used and a schematic of their disturbance geometries is shown in figure~\ref{fig:PertSchematic}. In figure~\ref{fig:obstacles}, the sub-panels on the left show the results of LDV measurements of the centre-line velocity (normalized by the bulk velocity $U$) $10 D$ downstream of the perturbation point. The corresponding sub-panels on the right indicate pressure traces  far downstream of the perturbation ($250 D$ and $2500 D$), showing typical puff sequences that the perturbation method generates. Figure~\ref{fig:obstacles}a shows measurements corresponding to the case when fluid is injected continuously through 20 holes in the pipe wall (figure~\ref{fig:PertSchematic}a). The measurements show (figure~\ref{fig:obstacles}a) that  although the flow profile is strongly distorted (the center-line velocity clearly deviates from 2U) and the velocity signal shows significant fluctuations, no puffs are triggered and the flow downstream is laminar. As already pointed out by \cite{hof2010eliminating}, continuous disturbances are inefficient in triggering puffs. By strongly flattening the profile the entire flow is turned refractory and turbulence collapses. Figure~\ref{fig:obstacles}b shows the results of placing a 2 mm orifice in the pipe (figure~\ref{fig:PertSchematic}b), and here occasionally single puffs emerge at irregular intervals. At slightly lower $Re$ ($\approx$ 2000) the orifice perturbation also stops creating puffs and only gives rise to laminar flow. The results of blocking the pipe with a semi-circular obstacle having a diameter of 4 mm (figure~\ref{fig:PertSchematic}c) are shown in figure~\ref{fig:obstacles}c, showing a larger density of puffs. The impulsive single-jet perturbation described earlier, on the other hand, readily triggers puffs each time it is actuated, provided that the perturbation amplitude is high enough and injections are spaced out by at least 25 advective time units to avoid strong interactions. Data from the impulsive jet (injection time of $\sim 2 D/U$ and mass flux$~10^{-2}$ of the pipe flow) are shown in figure~\ref{fig:obstacles}d, where puffs are triggered every 2 seconds. The same holds for the push-pull mechanism where fluid is injected and withdrawn simultaneously from diametrically opposite holes in the pipe. Here, a perturbation time of $\sim 20 D/U$ and a relative flux of $~10^{-3}$ was used. The length of the perturbation and the mass flux were chosen to be experimentally convenient, while ensuring that each perturbation event resulted in the production of precisely one puff. Due to the reproducible nature of impulsive perturbations (single impulsive jet and push-pull), they are used (with the perturbation times and fluxes mentioned here) for the remainder of the measurements. For the experiment with the impulsive jet, the pressure signal as the puff sequences pass the downstream sensor (at $2500 D$) (dotted line), reiterates the fact that, even at this Reynolds number (2100), puff sequences do not change much over the time available even in the longest laboratory pipes.

Though all puff sequences in figure~\ref{fig:obstacles} are recorded at the same $Re$, the different perturbation schemes give rise to very different flow patterns varying from high to sparse puff densities, and even fully laminar flow. From such measurements it is therefore impossible to define a critical point and the question arises if flows intrinsically depend on initial conditions or if flows on much longer time scales settle to a steady state, where the turbulent fraction (puff density) is uniquely defined by the Reynolds number. 



\section{Lifetime measurements}
\label{Lifetimes}

Dynamically, puffs are chaotic repellors \citep{Hof2008} and the chaotic dynamics make a precise prediction of the escape time from the chaotic set (as well as the occurrence of splitting events) impossible. Hence puffs evolve seemingly unaltered (regarding average quantities such as mean speed and size) on the chaotic set until they suddenly disappear. However, as has been shown previously, puffs have a characteristic lifetime which only depends on $Re$ \citep{Hof2008,DeLozar2009,avila2010transient,avila2011onset}. To test the different pipes used in this study and to provide a connection to previous work, we will in the following, determine puff lifetimes for the two different setups used in this study and compare them to previous experimental and numerical lifetime studies.

\begin{figure}
 \centerline{\includegraphics[scale=0.7]{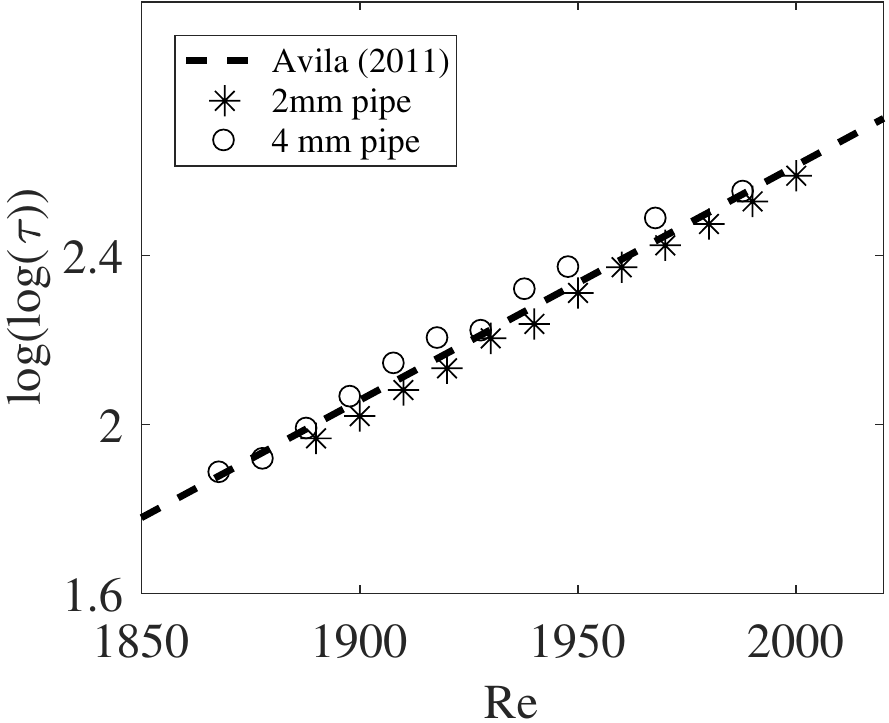}}%
\caption{Mean lifetime before puff decay in the 2 mm and 4 mm pipes compared with the non-diverging super-exponential scaling in  \cite{avila2011onset}(dashed line).}
 \label{lifetimes}
 \end{figure}

In the 2 mm pipe, puffs were generated one at a time, such that there was only one puff in the pipe at any instant. The pressure signal from the sensor at $7875 D$ was used to determine if puffs generated at the perturbation point survived to this location or had decayed upstream. For a given $Re$, 1000 puffs were investigated to obtain sufficient statistics for mean lifetimes. The survival probabilities of a puff are expected to follow $P(t)=exp(-(t-t_0)/\tau)$, where $\tau (Re)$ is the characteristic puff lifetime.  As shown in figure~\ref{lifetimes}, measured puff lifetimes are in excellent agreement with earlier lifetime studies \cite{Hof2008, DeLozar2009, Kuik2010a, avila2010transient, avila2011onset} which had been carried out in different pipe setups as well as obtained from direct numerical simulations. Measurements are repeated in the 4 mm pipe, where instead of a single jet, the push-pull injection (with zero net mass-flux) was used to create puffs. Again, lifetime statistics  are in excellent agreement with those of the 2 mm pipe and the studies cited above. 

Our data hence supports the superexponential lifetime scaling suggested previously and confirms that puff statistics only depend on $Re$ (not on specifics of a given pipe setup or the perturbation used). In particular, it rules out a transition scenario based on diverging lifetimes of single puffs.
It should be noted that the non-diverging super-exponential lifetime scaling only applies to single, isolated puffs. The lifetimes of a spatio-temporally intermittent flow, i.e. a flow comprising multiple interacting puffs, is expected to show diverging lifetimes (of turbulence overall) instead.  While individual puffs still decay, the creation of new puffs (puffs splitting) has been suggested \citep{avila2011onset} to eventually outweigh this process. Because of the extremely long time scales, the divergence of the overall lifetime could not be shown explicitly by \cite{avila2011onset}, instead this could only be inferred from decay and splitting rates of single puffs. Statistics of isolated puffs can also not give any insights on the eventual flow pattern, which is determined by puff interactions. The single puff statistics and the simple balance between puff splitting and decay rates predict a continuously increasing number of puffs above the critical point. With increasing turbulent fraction on the other hand puff-puff interactions will become important and are expected to saturate the growth process.

In the following, we will show that by exploiting the memoryless nature of puffs extremely long observation times (of the order of $10^8$ advective time units) can be accomplished which in turn allows the study of the long term evolution of the spatio-temporally intermittent flow pattern.

 \section{Steady state and critical point}
\label{Results}


 The central idea of the present study is to exploit this memoryless nature of puffs and the large separation between time scales, i.e. between the fast chaotic internal puff dynamics and the exceedingly slow evolution times of flow patterns. On the latter time scales, puffs have well defined average properties and are individually indistinguishable. 
We show that it is then possible to have periodic boundary conditions (in a sense that will be clarified shortly), that allows us to monitor puff sequences for longer than $10^7$ advective time units.

\begin{figure}

  \centerline{\includegraphics[scale=0.7]{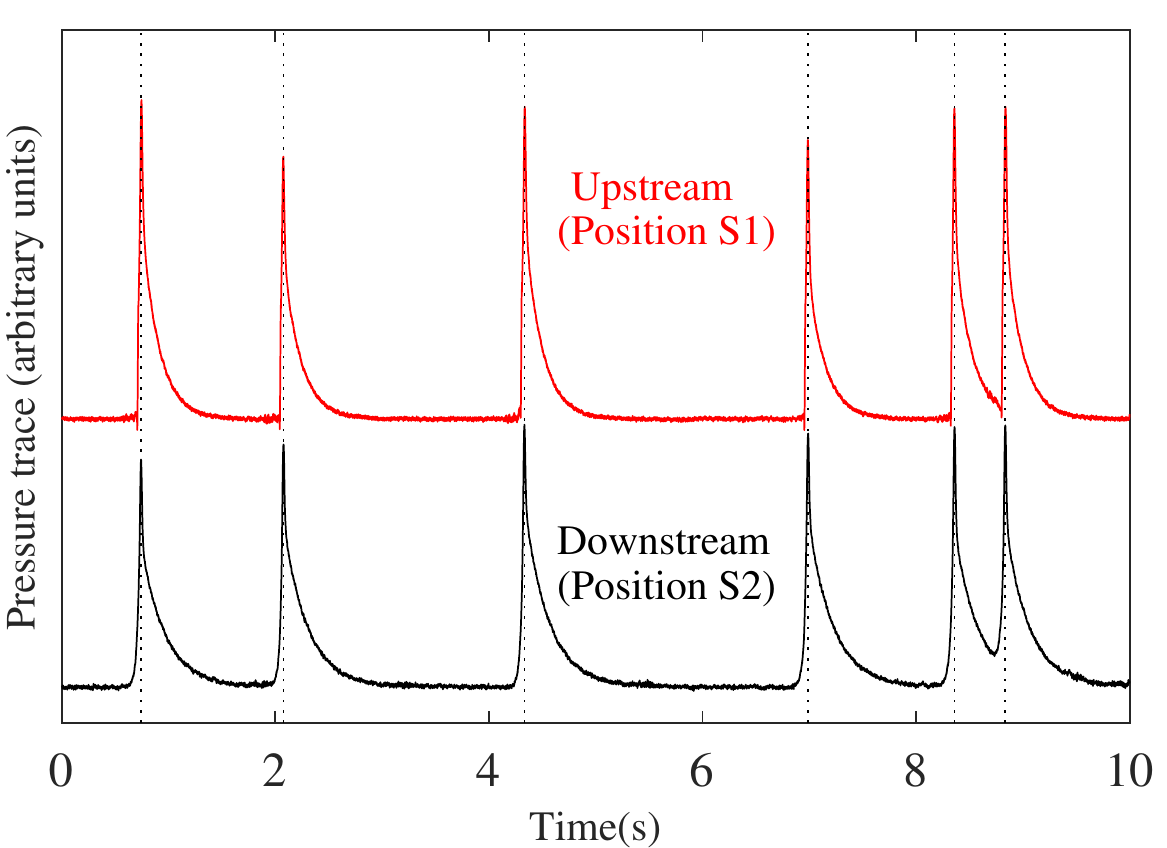}}
  \caption{Pressure traces of a puff sequence near the pipe exit (position S2 in figure~\ref{fig:schematic}) and that of the reproduced sequence near the entrance (position S1 in figure~\ref{fig:schematic})}.
\label{fig:puff reproduction}
\end{figure}

To do this, the flow is perturbed $150 D$ downstream of the inlet (point P in figure~\ref{fig:schematic}) to create a sequence of puffs that serves as the initial condition. This puff sequence travels downstream, and puffs can freely interact, split and decay. 
The resulting puff sequence is then monitored close to the pipe exit (position S2 in figure~\ref{fig:schematic}) by a pressure sensor. Each time a puff is detected at this point, a new puff is created at the perturbation point P. This way, the sequence of puffs leaving the pipe is recreated at the pipe entrance and the puff spacing is precisely kept as shown in figure~\ref{fig:puff reproduction}. If puffs are indeed memoryless and identical in all their mean properties as discussed in sections~\ref{puff characteristics} and  ~\ref{Lifetimes}, the recreated puff sequence  should in a statistical sense be equally representative and eventually lead to the same asymptotic state.  In this sense, the pipe setup provides periodic boundary conditions for the puff sequences studied and allows to monitor the evolution of these sequences for arbitrarily long times. If successful, this method should show if puff sequences eventually settle to a well defined statistical steady state with a fixed turbulent fraction. 

In the following, starting from some initial distribution of puffs, the exiting puff sequence will be reproduced at the inlet as described above. The evolution of the puff pattern is then monitored for long times to probe if eventually an equilibrium state is reached.

\begin{figure}
  \centerline{\includegraphics[scale=0.6]{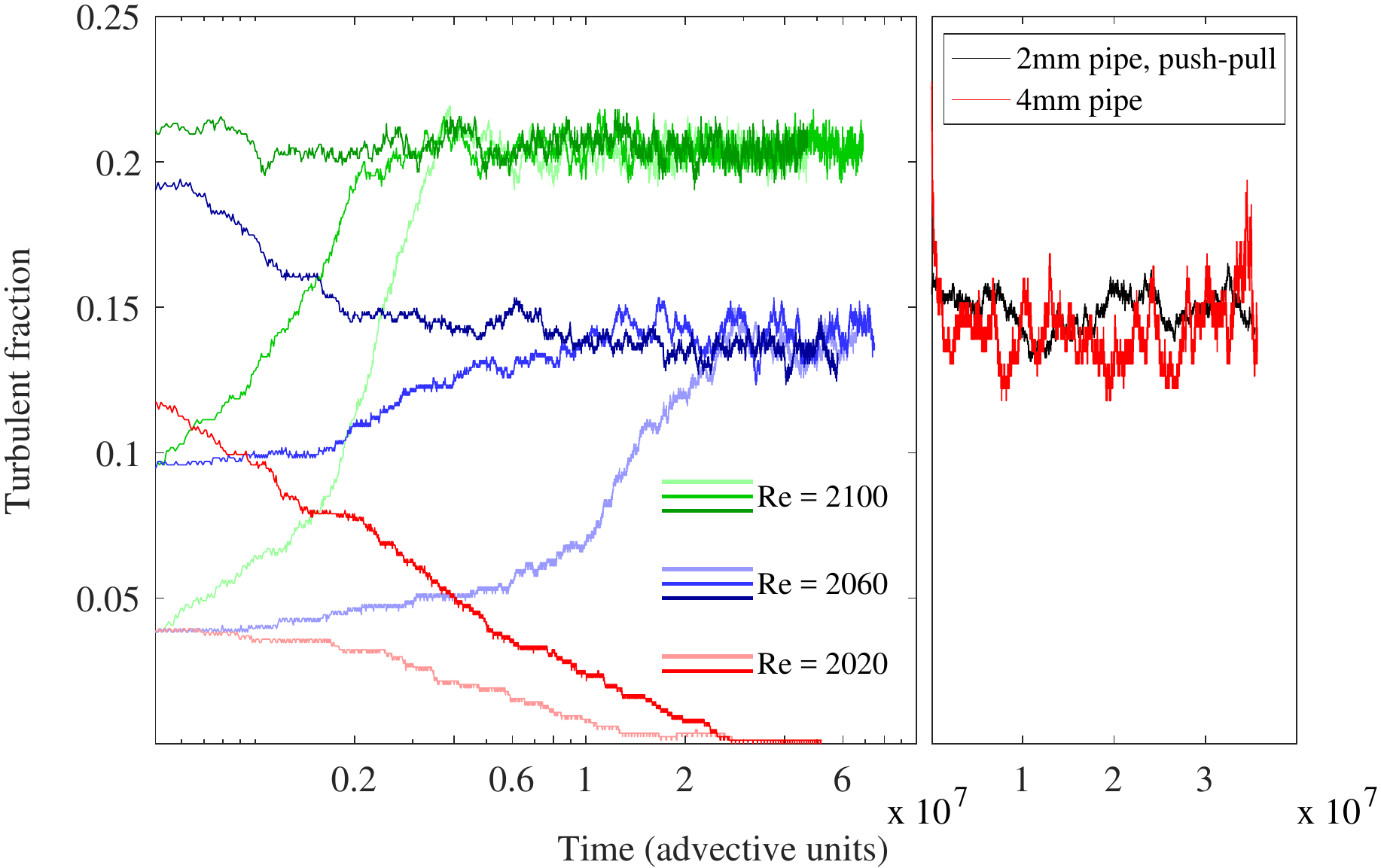}}
  \caption{Evolution of the turbulent fraction in the pipe: Left Panel: Results from the 2 mm pipe, with the impulsive jet used as a perturbation. Curves in red, blue and green correspond respectively to $Re =$ 2020, 2060 and 2100.
Right Panel: Result from the 2 mm pipe with the push-pull perturbation (black curve) and the 4 mm pipe with the impulsive jet (red curve), for $Re = 2060$. In the shorter 4 mm pipe, the temporal fluctuations are larger, because the turbulent fraction is averaged over a smaller system size.}
\label{fig:tf}
\end{figure}

\begin{figure}
 \centerline{\includegraphics[scale=0.5]{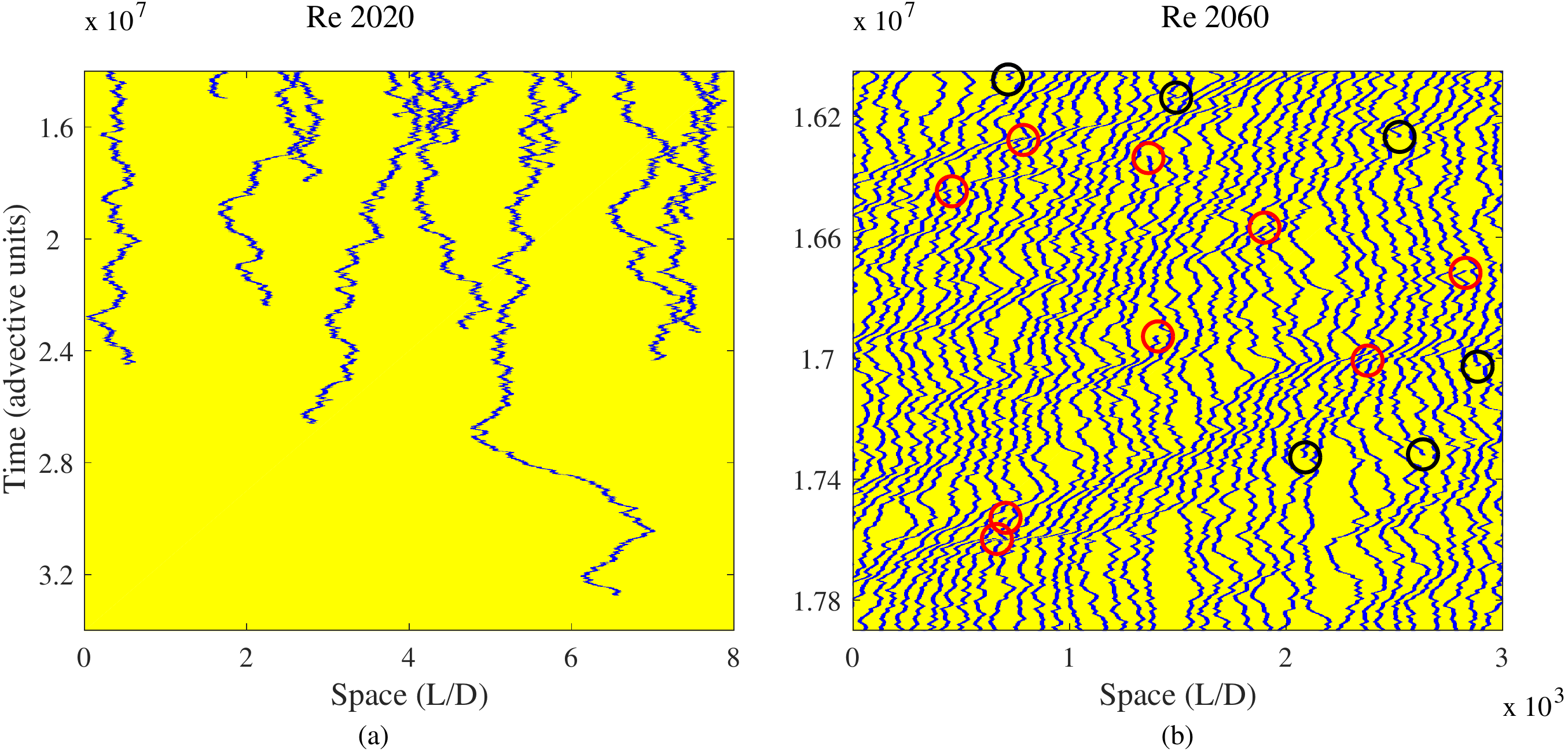}} 
 \caption{Space-time plots with the puffs shown in blue on a yellow laminar background. (a) Portion of a space-time plot for $Re = 2020$, where all the puffs eventually decay. (b) Portion of a space-time plot under a statistical steady-state condition for  $Re =2060$. Splitting and decay events are marked with red and black circles respectively. }
 \label{fig:SpaceTime}
\end{figure}

We first present the results of experiments in the $D = 2$ mm pipe, with the impulsive jet used as the perturbation. The left panel in figure~\ref{fig:tf} shows the time evolution of the turbulent fraction for different $Re$ and different initial conditions. Note that the time axis is in a logarithmic scale. Each color corresponds to a particular $Re$; the curves in red, blue and green correspond respectively to $Re =$ 2020, 2060 and 2100. The different shades of a color correspond to different initial (time t=0) turbulent fractions. The lighter and darker shades correspond respectively to lower and higher values of the initial turbulent fraction.
At first, we discuss the runs at $Re =$ 2020, which is 1$\%$ below the value of $Re_{c} = 2040$ proposed by \cite{avila2011onset}.  The initial condition chosen was a sequence of 32 equally spaced puffs. Taking the average length of the turbulent portion of a puff to be $5 D$, this corresponds to a turbulent fraction of 4$\%$. The length of the turbulent part of a puff depends to a certain extent on the criterion used to discriminate it from the surrounding laminar flow. However, a different choice only scales the turbulent fraction by a factor and does not change the conclusions obtained. Monitoring the flow at $Re = 2020$ starting from this initial condition, the puffs indeed eventually all decayed as seen by the light red curve in the left panel of figure~\ref{fig:tf}. Figure~\ref{fig:SpaceTime}a is a space-time plot showing the eventual return to laminar flow. It took more than $3 \times 10^{7}$ advective time units before the system settled to laminar flow and the longest surviving puff passed the pipe approximately 4000 times before it eventually decayed. The experiment was repeated for the same $Re$ but with a different initial condition (around 4 times larger turbulent fraction) and again the flow eventually completely relaminarized (dark red curve in figure~\ref{fig:tf}a).
 
In contrast to this at $Re =$ 2060, the flow settles to a statistical steady state where the turbulent fraction is non-zero. The measurements for this $Re$ are repeated for three different initial conditions: turbulent fractions of 4, 10 and 17$\%$ (corresponding respectively to 32, 80 and 136 puffs), the time evolution of which is indicated respectively by the light, medium and dark blue curves in figure~\ref{fig:tf}a. While the first two of these initial conditions had sequences of equally spaced puffs, the third one had a random puff sequence that was obtained downstream of a disturbed inlet (nozzle entrance) in a separate experiment.
In all the cases, the flow reaches a statistical steady state in which the turbulent fraction fluctuates about a well defined mean. The mean value of the turbulent fraction is the same for the different initial conditions, being around 14$\%$ (108 puffs). A portion of the space-time plot in the statistical steady state for $Re = 2060$ is shown in figure~\ref{fig:SpaceTime}b, with splitting and decay events indicated by red and black circles respectively. It should be emphasized that the existence of an equilibrium level of the turbulent fraction is a direct consequence of spatial interactions and cannot be predicted from single puff statistics. 

Similarly, for $Re =$ 2100, the evolution for initial conditions corresponding to turbulent fractions of 4, 10 and 22 $\%$ are shown by the light, medium and dark green curves respectively. For this $Re$, the steady state corresponds to a turbulent fraction of around 21 $\%$. The time taken to reach steady state is smaller while the frequency of fluctuations is higher as compared to $Re =$ 2060, as would be expected because puff splittings occur more frequently at this higher $Re$.

Hence we first observe sustained `puff-turbulence' at $Re=2060$ and the critical point therefore falls in between $2020<Re_c<2060$  showing that the estimate obtained by a simple balance of decay and splitting times \citep{avila2011onset}  works well for the case of pipe flow. 
To test if the determined steady-state turbulent fraction depends on the pipe setup or the type of perturbation used, the experiment was repeated for $Re =$ 2060 in the 2 mm pipe, using the push-pull perturbation and subsequently the pipe was changed to the 4 mm pipe. In the latter the impulsive jet perturbation was used. As discussed in section 2, apart from having a greater diameter, this pipe has a shorter length and a lower $Re$ for the natural transition. The results from these runs are shown in figure~\ref{fig:tf}b and in both cases the turbulent fraction fluctuates around the same mean as in the original 2 mm pipe experiments figure~\ref{fig:tf}a. It hence appears that the average turbulent fraction is independent of the pipe and only a function of $Re$.

\begin{figure}
  \centerline{\includegraphics[scale=0.75]{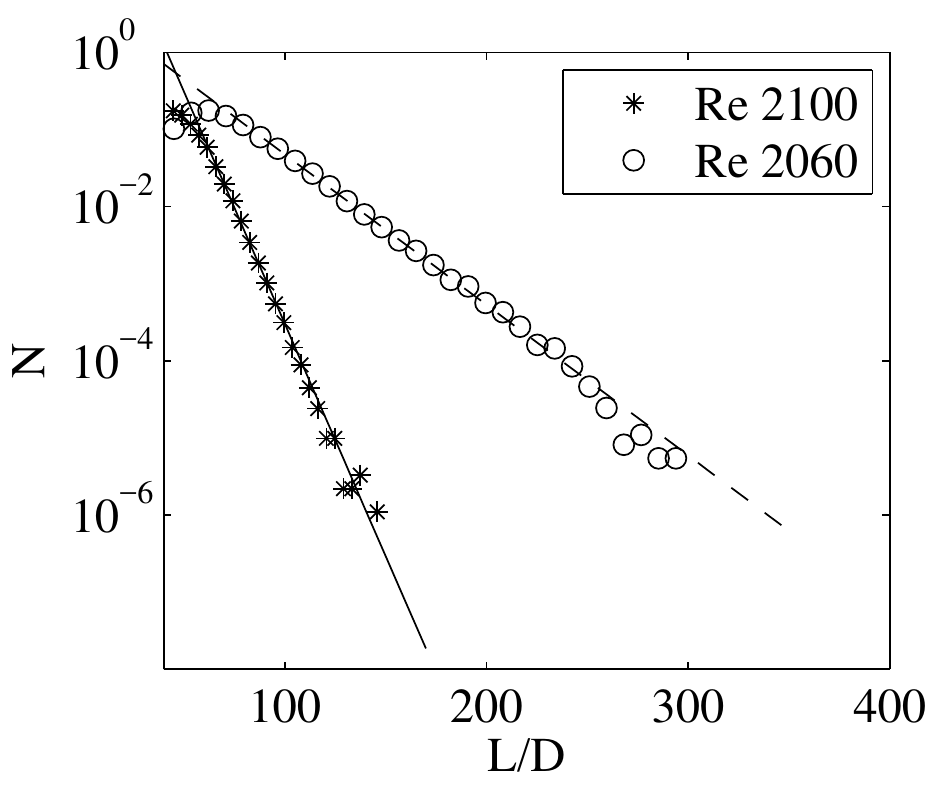}}
  \caption{Probability distributions of laminar gap sizes at $Re =2060$ and $2100$}
\label{fig:dist}
\end{figure}

It is known that certain spatio-temporal systems exhibit scale invariant flow patterns close to the critical point \citep{lemoult2016directed,chate1988spatio}. In such cases, the size distribution of laminar gaps follows a power-law distribution close enough to the critical point, and exponential distributions farther away. We computed the distributions in the steady state for $Re =$ 2060 and 2100, and these are shown in figure~\ref{fig:dist}. For both these Reynolds numbers, we find that the laminar gap sizes are
exponentially distributed. According to the measurements discussed above, the critical point lies between $Re = 2060$ and $Re = 2020$. Hence, our measurements could be as much as 2$\%$ away from the critical point, which could be insufficient to observe scale invariant behaviour. In the present setup, a more accurate determination of the critical points and measurements even closer to it are not possible due to limitations in the long-term stability of the Reynolds number
(0.25$\%$). The local dynamics of puffs and their interactions also play a role in deciding how close to the critical point, if at all, scale-free behaviour is observed, and some aspects are discussed in the next section.

\section{Spatio-temporal intermittency and puff clustering}

Above the critical point, the flow consists of turbulent puffs separated by laminar regions of varying length and the overall pattern continuously changes in time as shown in figure~\ref{fig:SpaceTime}. Spatio-temporal intermittency is hence an intrinsic feature of flows in this regime. Overall the spatio-temporal intermittent nature of the flow pattern that is assumed in the long time limit is characteristic for spreading processes in out-of-equilibrium (i.e. dissipative) systems  where a fluctuating state (here turbulence) competes with an absorbing state (here laminar flow). In pipe flow, the laminar state qualifies as an absorbing state because of its linear stability, i.e. if a region relaminarizes it remains in the laminar state for all times unless perturbed by a neighbouring turbulent site. If all turbulence has decayed, the flow subsequently remains laminar. The analogy between linearly stable shear flows and non-equilibrium phase transitions from fluctuating into absorbing states has first been pointed out by \cite{pomeau1986front}. Subsequently, the physical processes that give rise to the competition between laminar and turbulent flow regions have been identified for pipe flow \citep{avila2011onset}. Here, spreading of turbulence occurs in the form of puff splitting, and the transient nature \citep{Hof2006a} of puffs and interactions \citep{hof2010eliminating} give rise to decay and relaminarization. Generally, stochastic spreading phenomena have been studied in great detail in statistical physics. Examples of absorbing state transitions \citep{hinrichsen2000non}  are epidemics (here healthy individuals can only be infected by contact to infected ones) or water `percolating' through a porous medium (here dry sites can only become wet if they are in direct contact with wet sites). With increasing control parameter, these systems display a continuous phase transition with universal features. They belong to the so called `directed percolation' (DP) universality class and are characterised by three independent critical exponents: the first predicts the density of active sites (i.e in our case the puff density or turbulent fraction) as the critical point is surpassed. The second and third exponents determine the spatial and temporal correlations of the statistical steady state close to the critical point; in our case, these would be the typical size of the laminar gaps along the space and time directions in a space-time plot such as figure~\ref{fig:SpaceTime}b. The DP framework hence predicts the characteristics of the spatio-temporal intermittent patterns that emerge at the phase transition but it does not make any predictions about the `microscopic' details, i.e. in the present case the chaotic internal dynamics of puffs and their stochastic interactions. Or in other words, the universal properties that emerge at the critical point are independent of the details of the microscopic dynamics \citep{hinrichsen2000non}. In recent experiments and numerical simulations, transition in Couette flow was shown to fall into the directed percolation universality class \citep{lemoult2016directed}. Other recent experiments in Poiseuille flow also proposed to find evidence for directed percolation \citep{sano2015universal}. In the latter case however, the critical point does not agree with the values found in recent studies (\cite{xiong2015turbulent}, Tsukahara (private communications) and Mukund, Parnajape and Hof (in preparation)). Moreover the equilibration times proposed by Sano and Tamai are approximately 5 orders of magnitude shorter than the splitting and decay times observed in pipes \citep{avila2011onset}. The finding of our present study that equilibration times exceed the spreading and decay times suggests that the flows investigated by \citep{sano2015universal} were far from reaching a statistical steady state.

While a connection between transition in pipe flow and the simple percolation framework may appear far fetched, DP is conjectured to universally apply to seemingly unrelated systems if the following four conditions are met    \citep{janssen1981nonequilibrium,grassberger1982phase,hinrichsen2000non}:
(i) the system undergoes a continuous phase transition from a fluctuating (turbulence) into a unique absorbing state (laminar flow), (ii) the system is characterized by a positive one component order parameter (in our case, the turbulent fraction), (iii) there are no long range interactions, and (iv) absence of additional symmetries or quenched randomness. While pipe flow appears to meet conditions 1, 2 and 4, we will in the following take a closer look at puff interactions.

\begin{figure}
  \centerline{\includegraphics[scale=0.6]{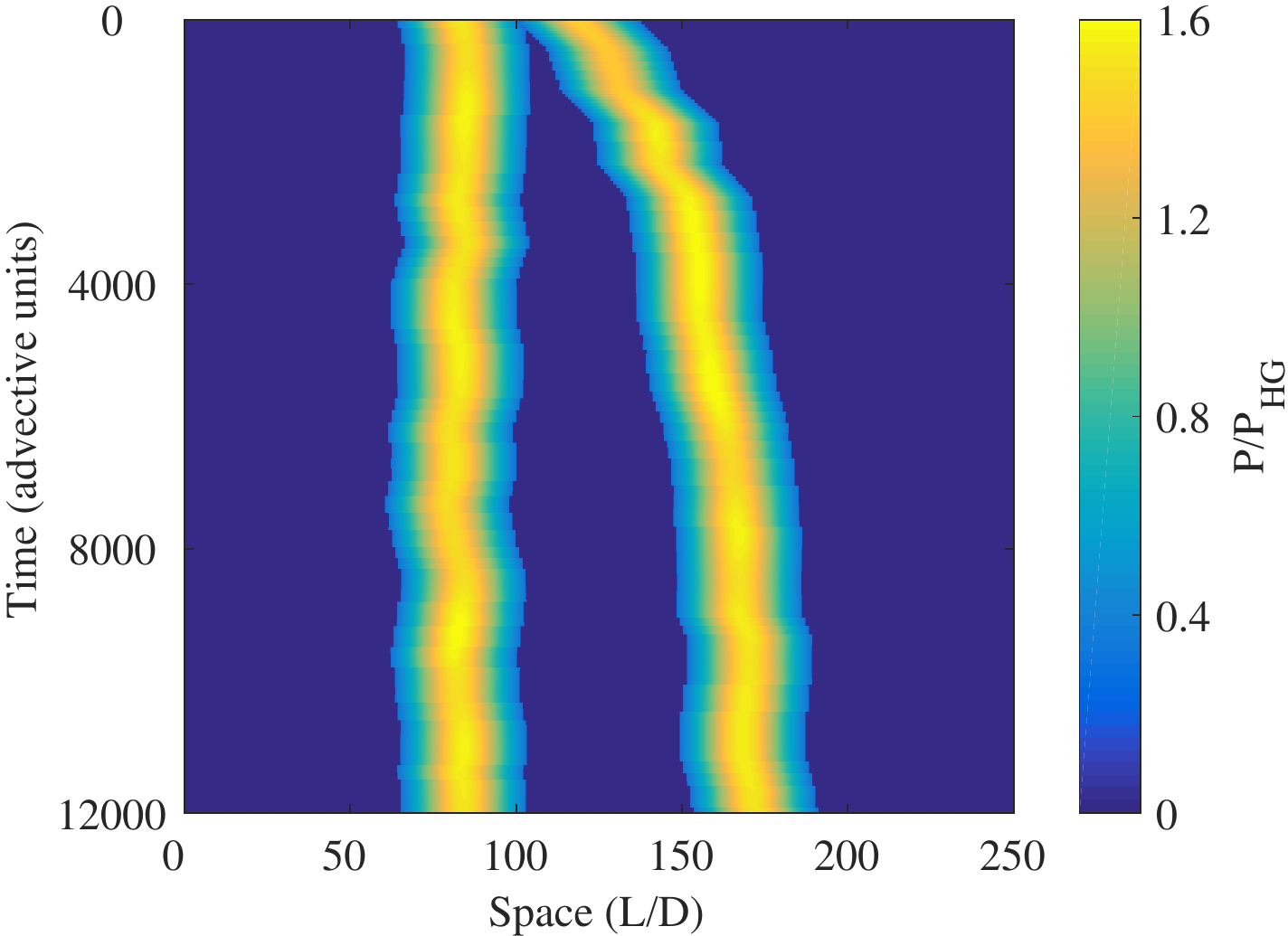}}
  \caption{A space-time plot of pressure showing the relative motion of two puffs that start with an initial separation of $30 D$. The pressure is measured across $8 D$ and normalized by the laminar pressure drop ($P_{HG}$) across the same distance. }
\label{fig:PuffRepulsion}
\end{figure}

A first consequence of puff interactions is that puff decay rates differ strongly from those for individual puffs measured by \cite{avila2011onset}. For the  approximately 40  puffs shown in figure~\ref{fig:SpaceTime}, single puff statistics would predict at most one decay to occur over the $2 \times 10^6$ time units displayed. However six decays are observed. In general, the relatively dense puff pattern considerably limits the expansion rates and increases puff decay rates (puffs in too close proximity can annihilate each other \citep{samanta2011experimental}. Overall these puff-puff interactions lead to a balance between splitting and decay events and hence are responsible for the flow to assume a steady turbulent fraction. Also, the time taken to reach equilibrium exceeds the time scales of splitting and decay at the critical point \citep{avila2011onset}, though the critical point has been approached to only within 1$\%$. 
Puff interactions are also directly visible in the space-time plot (figure~\ref{fig:SpaceTime}) where puff clusters are observed, which propagate in the streamwise direction (diagonal lines going from top right to bottom left). To better understand the cause of puff clustering we carried out a separate experiment where two puffs are generated $30 D$ apart and monitored with pressure sensors every $500 D$. Five independent runs were carried out and the pressure signal as well as the arrival times at each of the sensor locations was averaged over these 5 independent puff pairs. Pressure values and arrival times at locations in between two adjacent sensors was linearly interpolated. The results are shown in a spacetime plot figure~\ref{fig:PuffRepulsion}, in a frame that is moving with the mean speed of a single, isolated puff. The mean speed of an isolated puff was measured in a separate experiment and its $Re$ dependence was found to be well described by $U_{p} = U(6.22 \times 10^{-7} Re^{2} - 2.83 \times 10^{-3} Re + 4.16)$ in the range $1700 < Re < 2250$, where $U_{p}$ is the speed of the puff and $U$ is the bulk flow speed. This  agrees well with previous measurements \citep{DeLozar2009,barkley2015rise}. Figure~\ref{fig:PuffRepulsion} shows that the mean velocity of the upstream puff is the same as that of an isolated puff. At separations less than around $100 D$, the downstream puff is affected by the upstream one and energetically weakened  \citep{hof2010eliminating}, as seen in a lower pressure drop across the puff. The weaker downstream puff then travels faster and the two puffs separate till they are  around $100 D$ apart; see also figure 27 in \cite{barkley2016theoretical}. Returning to figure~\ref{fig:SpaceTime}, puff clustering is generally caused by puffs being in close range to their upstream neighbours. The energetically weakened puff travels faster while the speed of the upstream puff is not affected (also see figure~\ref{fig:PuffRepulsion}). The faster puff then separates from its upstream neighbour so that it eventually slows down again, exactly as shown for the two interacting puffs in figure~\ref{fig:PuffRepulsion}. However at the same time it gets into range of its next downstream neighbour which in turn accelerates. The resulting cluster of puffs with increased speed then propagates down the pipe in a wave like fashion. This clustering breaks the left-right symmetry in the space-time plot whose visual appearance differs from space-time plot for directed percolation. This could potentially have consequences for the nature of the transition.

Puff clustering clearly increases the interaction length beyond the nearest neighbour limit. It typically involves a group of around 5 puffs. Whether the interaction stops at this range (i.e. falls off exponentially) or if a weak interaction persists to much greater length (in the extreme case the entire pipe) is unclear. In the former case, the transition is likely to still fall into the DP universality class but would potentially require larger system sizes to determine the nature of the transition. Such interactions also have an influence on the range of $Re$ near the critical point where the universal scaling applies. The lack of scale invariance at $Re = 2060$ suggests that the scaling range is considerably smaller than $2\%$ in $Re$. The possible requirement of even larger system sizes and tighter control on the long term stability of $Re$ pose considerable challenges for future experiments. In any case, pipe flow appears to be more complicated than Couette flow where such interactions due to the upstream-downstream symmetry are absent.

%
\section{Conclusion}
We show that spatio-temporally intermittent flow patterns in pipe flow settle to a well defined statistical steady state with a finite turbulent fraction that is uniquely defined by the Reynolds number. Puff-puff interactions dictate the equilibrium state that is asymptotically approached: a state of persistent intermittency where the ratio of turbulent to laminar regions has reached a fixed value. Interactions between closely spaced puffs result in clustering of puffs that appears to be unique to pipe flow and could potentially alter the nature of the transition or the system size required to characterize it.
On the other hand, puff interactions do not appear to have a strong effect on the value of the critical point (less than 1$\%$). The method used in the present study in principle can be applied even closer to the critical point to elucidate further details of the transition.

The extremely long equilibration times encountered explain the wide scatter of the values of the critical point reported over the last 130 years. Observation times in earlier studies were typically more than four orders of magnitude shorter than the time scales needed  to determine an accurate value for the onset of sustained turbulence. In such insufficiently long pipes, the turbulent fraction observed is then entirely dictated by the disturbance levels responsible for creating turbulence at the pipe inlet (which differs from experiment to experiment). We argue that similarly in other shear flows like duct, channel and Couette flow the lack of agreement on a critical point and the nature of the transition is at least partially caused by limitations in observation times. The approach used in the present study can be equally applied to other shear flows and should eventually enable to characterize the nature of the transition in different flows.

\bibliographystyle{jfm}

\bibliography{periodic_pipe}

\end{document}